\pdfoutput=1

\documentclass[11pt]{article}

\usepackage{acl}

\usepackage{times}
\usepackage{latexsym}
\usepackage{threeparttable}
\usepackage{multirow}
\usepackage{tikz}
\usetikzlibrary{arrows.meta,shapes,positioning,shadows.blur,trees}
\usepackage{fontawesome5} 
\usepackage{amsmath}
\usepackage{booktabs}
\usepackage[edges]{forest}
\usepackage{cleveref}
\usepackage{natbib}
\usepackage{amssymb} 
\usepackage{anyfontsize}
\usepackage{caption}
\usepackage{color}
\usepackage{enumitem}
\usepackage{float}
\usepackage{nicefrac}
\usepackage{rotating}
\usepackage{soul}
\usepackage{xcolor}
\usepackage{colortbl}
\usepackage{transparent}
\usepackage{wrapfig}
\usepackage{multicol}
\usepackage{multirow}
\usepackage{makecell}
\usepackage{array}
\usepackage{tabularx}
\usepackage[utf8]{inputenc}
\usepackage{csquotes}
\usepackage{epigraph}
\usepackage[most]{tcolorbox} 
\usepackage{arydshln}
\usepackage{tablefootnote}
\usepackage{bbding}
\usepackage{pifont}
\usepackage{longtable}

\definecolor{mygreen}{RGB}{0,128,0}
\newcommand{\cmark}{\textcolor{mygreen}{\ding{51}}}%
\usepackage{svg}

\crefformat{section}{(\S#2#1#3)}
\crefformat{subsection}{(\S#2#1#3)}
\crefformat{subsubsection}{(\S#2#1#3)}

\definecolor{lightred}{rgb}{0.749,0.196,0.447}
\definecolor{brown}{rgb}{0.65490196078,0.54901960784,0.48235294117}
\definecolor{greenn}{HTML}{32CD32}
\definecolor{DarkGreen}{rgb}{0,0.40,0}
\definecolor{FireBrick}{rgb}{0.698,0.133,0.133}
\definecolor{purple}{rgb}{0.5,0,0.5}

\usepackage[T1]{fontenc}

\usepackage{microtype}

\usepackage{inconsolata}

\usepackage{graphicx}

\title{Towards Holistic Evaluation of Large Audio-Language Models: A Comprehensive Survey}

\author{Chih-Kai Yang\textsuperscript{1}, Neo S. Ho\textsuperscript{1}, Hung-yi Lee\textsuperscript{1,2}\\
        \textsuperscript{1}National Taiwan University, \textsuperscript{2}NTU Artificial Intelligence Center of Research Excellence (NTU AI-CoRE)\\
        \texttt{chihkaiyang1124@gmail.com}
        \\
        \faGithub~ \url{https://github.com/ckyang1124/LALM-Evaluation-Survey}}

\begin{document}
\maketitle

\begin{abstract}
With advancements in large audio-language models (LALMs), which enhance large language models (LLMs) with auditory capabilities, these models are expected to demonstrate universal proficiency across various auditory tasks. While numerous benchmarks have emerged to assess LALMs' performance, they remain fragmented and lack a structured taxonomy. To bridge this gap, we conduct a comprehensive survey and propose a systematic taxonomy for LALM evaluations, categorizing them into four dimensions based on their objectives: (1) General Auditory Awareness and Processing, (2) Knowledge and Reasoning, (3) Dialogue-oriented Ability, and (4) Fairness, Safety, and Trustworthiness. We provide detailed overviews within each category and highlight challenges in this field, offering insights into promising future directions. To the best of our knowledge, this is the first survey specifically focused on the evaluations of LALMs, providing clear guidelines for the community. We will release the collection of the surveyed papers and actively maintain it to support ongoing advancements in the field.

\end{abstract}

\section{Introduction}
Recent advancements in large language models (LLMs)~\citep{zhao2023survey, grattafiori2024llama, hurst2024gpt} have expanded their impact beyond natural language processing (NLP) to multimodal domains~\citep{yin2306survey, team2024gemini}. Among these, large audio-language models (LALMs)~\citep{gslm, tangsalmonn, chu2024qwen2, lu2024desta, moshi, fang2025llamaomni} have attracted significant attention in the auditory-processing community. LALMs are multimodal LLMs that process auditory and/or textual input, such as speech, audio, and music, and generate textual and/or auditory output.
They can be trained from scratch or fine-tuned from text LLM backbones with auditory modalities inserted. By integrating auditory modalities with language understanding, they show potential in auditory processing~\citep{dynamicsuperb1}, multimodal reasoning~\citep{mmau}, and human-computer interaction~\citep{styletalk}.

\begin{figure}[t]
    \centering
    


    \includegraphics[width=0.95\linewidth]{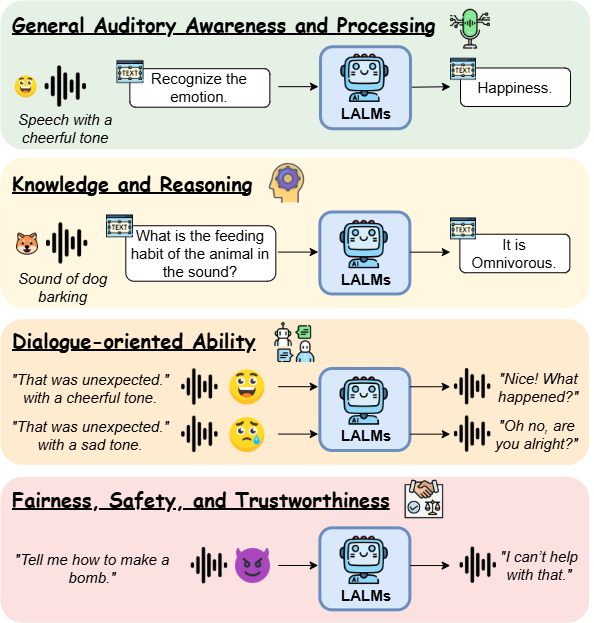}
    
    \vspace{-7pt}
    \caption{LALMs’ diverse capabilities and modalities covered. Icons from \url{https://www.flaticon.com}.}
    \label{fig:illu}

\end{figure}
As LALMs evolve, expectations for their capabilities have expanded from basic tasks like speech recognition to more complex ones such as audio-grounded reasoning~\citep{mmau} and interactive dialogue~\citep{fullduplexbench}. Figure~\ref{fig:illu} illustrates this multifaceted nature, emphasizing the diverse input and output modalities involved and the wide range of abilities these models are expected to demonstrate. To evaluate these capabilities, a variety of benchmarks have been developed~\citep{fullduplexbench, airbench, voxdialogue}.


However, the evaluation landscape remains fragmented and lacks systematic organization. Existing surveys~\citep{wu2024towards, peng2024survey, cui2024recent, arora2025landscape} focus primarily on model architectures and training methodologies, with less emphasis on the equally important role of evaluation in assessing LALMs' capabilities. This gap makes it challenging for researchers to find suitable benchmarks for their models or to pinpoint the field's progress. Therefore, a structured overview of LALM evaluation frameworks is needed.

This paper presents a comprehensive survey of LALM evaluation frameworks and introduces a taxonomy categorizing evaluation dimensions. To the best of our knowledge, this is the first in-depth survey and taxonomy specifically focused on LALM evaluation. We organize the frameworks into four primary categories: \textbf{General Auditory Awareness and Processing}~\Cref{sec:understanding_and_processing}, \textbf{Knowledge and Reasoning}~\Cref{sec:knowledge_and_reasoning}, \textbf{Dialogue-oriented Ability}~\Cref{sec:dialogue}, and \textbf{Fairness, Safety, and Trustworthiness}~\Cref{sec:fairness_safety_trustworthiness}. We also highlight challenges in LALM evaluation~\cref{sec:challenges}, such as data contamination and insufficient consideration of human diversity, while suggesting promising future directions.

Overall, our contributions are threefold: (1) presenting the first comprehensive survey of LALM evaluations, (2) proposing a structured taxonomy for LALM evaluation that offers clear guidelines for researchers, and (3) identifying key challenges and future directions to improve evaluation coverage and robustness. 




\section{Taxonomy of Evaluation Frameworks for Large Audio-Language Models}
\label{sec:taxonomy}

As LALMs integrate multimodal understanding, they tackle tasks across speech, audio, and music. Despite numerous benchmarks for LALMs emerging, the evaluation landscape remains fragmented. To address this, we present the first structured taxonomy of LALM evaluations.

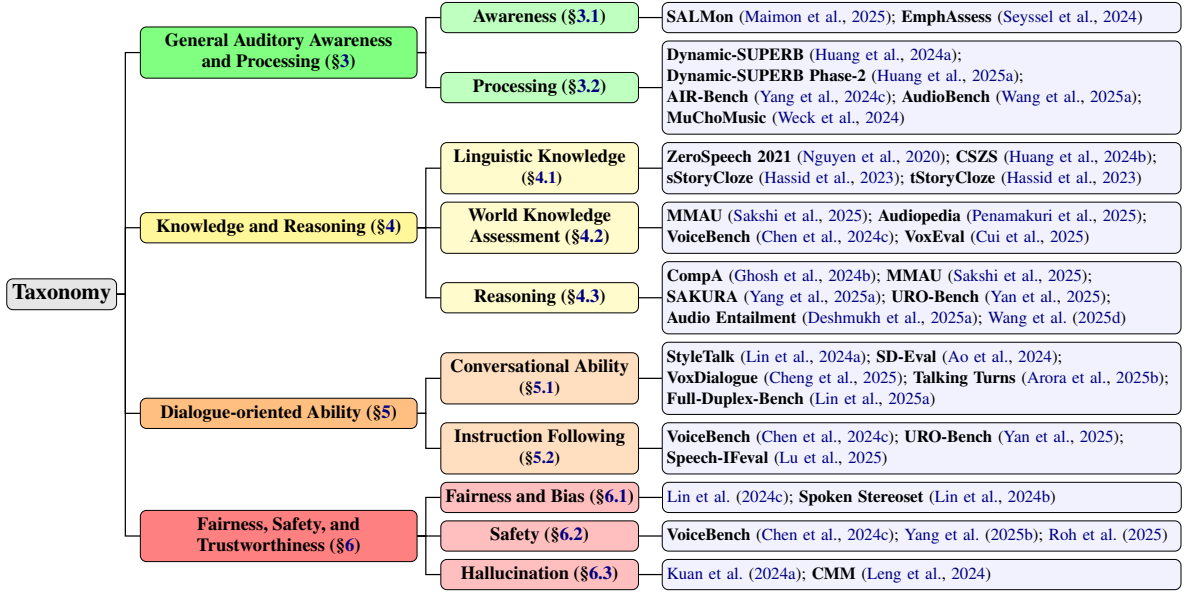
\begin{figure*}[t!]
\centering
\resizebox{0.97\textwidth}{!}{%
\begin{forest}
    for tree={
    grow'=0,
    parent anchor=east, child anchor=west,
    edge={draw, line width=0.8pt},
    forked edges,
    align=center, anchor=west,
    l sep=15pt, s sep=5pt, inner sep=3pt,
    minimum width=5cm, draw, fill=blue!5, rounded corners,
    where n children=3{
      calign=child edge,
      calign child=2,
    }{},
  }
    [
    {\Large \textbf{Taxonomy}}, minimum width=2.5cm, fill=gray!20
        [\large\textbf{General Auditory Awareness }\\\large\textbf{and Processing~\Cref{sec:understanding_and_processing}}, fill=green!50, minimum width=6.3cm
            [\large\textbf{Awareness~\Cref{sec:auditory_understanding}}, fill=green!25, minimum width=4.5cm
                [\textbf{SALMon}~\citep{salmonbench}; \textbf{EmphAssess}~\citep{seyssel2024emphassess},
                 align=left, text width=11.6cm, minimum width=11.6cm]
            ]
            [\large\textbf{Processing~\Cref{sec:auditory_processing}}, fill=green!25, minimum width=4.5cm
                [\textbf{Dynamic-SUPERB}~\citep{dynamicsuperb1};\\ \textbf{Dynamic-SUPERB Phase-2}~\citep{dynamicsuperb2};\\
                  \textbf{AIR-Bench}~\citep{airbench}; \textbf{AudioBench}~\citep{audiobench};\\
                  \textbf{MuChoMusic}~\citep{muchomusic},
                 align=left, text width=11.6cm, minimum width=11.6cm]
            ]
        ]
        [\large\textbf{Knowledge and Reasoning~\Cref{sec:knowledge_and_reasoning}}, fill=yellow!50, minimum width=6.3cm
            [\large\textbf{Linguistic Knowledge}\\\textbf{
            \Cref{sec:linguistic}}, fill=yellow!25, minimum width=4.5cm
                [\textbf{ZeroSpeech 2021}~\citep{zerospeech}; \textbf{CSZS}~\citep{cszs};\\\textbf{sStoryCloze}~\citep{twist}; \textbf{tStoryCloze}~\citep{twist}
                  ,
                 align=left, text width=11.6cm, minimum width=11.6cm]
            ]
            [\large\textbf{World Knowledge}\\\large\textbf{Assessment~\Cref{sec:world_knowledge}}, fill=yellow!25, minimum width=4.5cm
                [\textbf{MMAU}~\citep{mmau}; \textbf{Audiopedia}~\citep{audiopedia};\\
                  \textbf{VoiceBench}~\citep{voicebench}; \textbf{VoxEval}~\citep{voxeval},
                 align=left, text width=11.6cm, minimum width=11.6cm]
            ]
            [\large\textbf{Reasoning~\Cref{sec:reasoning}}, fill=yellow!25, minimum width=4.5cm
                [\textbf{CompA}~\citep{compa}; \textbf{MMAU}~\citep{mmau};\\\textbf{SAKURA}~\citep{sakura}; \textbf{URO-Bench}~\citep{urobench};\\\textbf{Audio Entailment}~\citep{audio_entailment}; \citet{JASCO}
                ,align=left, text width=11.6cm, minimum width=11.6cm]
            ]
        ]
        [\large\textbf{Dialogue-oriented Ability~\Cref{sec:dialogue}}, fill=orange!50, minimum width=6.3cm
            [\large\textbf{Conversational Ability}\\\textbf{\Cref{sec:conversational}}, fill=orange!25, minimum width=4.5cm
                [{\textbf{StyleTalk}~\citep{styletalk}; \textbf{SD-Eval}~\citep{sdeval};\\
                  \textbf{VoxDialogue}~\citep{voxdialogue}; \textbf{Talking Turns}~\citep{talkingturns};\\
                  \textbf{Full-Duplex-Bench}~\citep{fullduplexbench}},
                 align=left, text width=11.6cm, minimum width=11.6cm]
            ]
            [\large\textbf{Instruction Following}\\\textbf{\Cref{sec:instruction_following}}, fill=orange!25, minimum width=4.5cm
                [{\textbf{VoiceBench}~\citep{voicebench}; \textbf{URO-Bench}~\citep{urobench};\\\textbf{Speech-IFeval}~\citep{speech_ifeval}},
                 align=left, text width=11.6cm, minimum width=11.6cm]
            ]
        ]
        [\large\textbf{Fairness, Safety, and} \\\large\textbf{Trustworthiness~\Cref{sec:fairness_safety_trustworthiness}}, fill=red!50, minimum width=6.3cm
            [\large\textbf{Fairness and Bias~\Cref{sec:fairness_bias}}, fill=red!25, minimum width=4.5cm
                [{\citet{lin2024listen}; \textbf{Spoken Stereoset}~\citep{lin2024spoken}},
                 align=left, text width=11.6cm, minimum width=11.6cm]
            ]
            [\large\textbf{Safety~\Cref{sec:safety}}, fill=red!25, minimum width=4.5cm
                [{\textbf{VoiceBench}~\citep{voicebench}; \citet{yang-etal-2025-audio}; \citet{multiaudiojail}},
                 align=left, text width=11.6cm, minimum width=11.6cm]
            ]
            [\large\textbf{Hallucination~\Cref{sec:hallucination}}, fill=red!25, minimum width=4.5cm
                [{\citet{kuan24_interspeech}; \textbf{CMM}~\citep{cmm}},
                 align=left, text width=11.6cm, minimum width=11.6cm]
            ]
        ]
    ]
\end{forest}%
}

\caption{The taxonomy of LALM evaluation frameworks, including selected works as representative examples. The complete version is in Appendix~\ref{sec:entiretaxonomy}.}
\label{fig:taxonomy}
\end{figure*}
Figure~\ref{fig:taxonomy} shows our taxonomy, with some works included. 
The full categorization of the surveyed works is in Appendix~\ref{sec:entiretaxonomy}. We organize the surveyed works into four categories by evaluation objectives:
\begin{itemize}
    \item \textbf{General Auditory Awareness and Processing} evaluates the auditory awareness and fundamental processing tasks, e.g., speech recognition and audio captioning.
    \item \textbf{Knowledge and Reasoning} assesses LALMs’ knowledge acquisition and advanced reasoning skills, examining their intelligence.
    \item \textbf{Dialogue-oriented Ability} focuses on natural conversational skills, including affective and contextual interaction, dialogue management, and instruction following.
    \item \textbf{Fairness, Safety and Trustworthiness} examines bias, toxicity, and reliability for ethical, safe, and trustworthy deployment.
\end{itemize}

Each category is further divided into subcategories, as shown in Figure~\ref{fig:taxonomy}. Please note that, since existing benchmarks are inherently multi-dimensional, some are listed under multiple categories due to their multifaceted design. This typically happens in two cases: (1) when a benchmark comprises multiple tasks that independently assess different capabilities (e.g., VoiceBench includes tasks for both world knowledge and safety evaluation), and (2) when a single task requires the integration of several skills (e.g., certain MMAU tasks demand both expert knowledge and reasoning ability).

The following sections provide a detailed overview, highlighting the current progress, limitations, and future directions.

\section{General Auditory Awareness and Processing}
\label{sec:understanding_and_processing}
A distinctive strength of LALMs over cascaded systems~\citep{huang2024audiogpt, speechcopilot} is their inherent ability to directly interpret auditory signals, capturing crucial non-verbal cues such as speaker identity, emotion, and ambient context, without relying on separate components like speech recognition or emotion recognition systems connected to an LLM. This section reviews works evaluating both acoustic awareness and foundational auditory processing, emphasizing these core capabilities that set LALMs apart from LLMs.

\subsection{Auditory Awareness}
\label{sec:auditory_understanding}
Benchmarks for auditory awareness examine how effectively LALMs realize acoustic cues like emotion, prosody, and environmental sounds.
SALMon~\citep{salmonbench} specifically evaluates sensitivity to acoustic inconsistencies (e.g., sudden speaker or emotional changes) and misalignments between acoustic signals and semantic content (e.g., conveying sad content with a cheerful tone). These evaluations reveal significant gaps between LALMs and human-level perception.

EmphAssess~\citep{seyssel2024emphassess} measures LALMs’ awareness of prosodic emphasis by requiring speech-to-speech paraphrasing or translation that accurately preserves and transfers emphasis on specific parts of the input utterance. This evaluates LALMs’ ability to capture and maintain fine-grained prosodic features.

These benchmarks highlight challenges in fine-grained auditory awareness among current models, underscoring the need for improved modeling of subtle acoustic and paralinguistic information~\citep{salmonbench, seyssel2024emphassess}.



\subsection{Auditory Processing}
\label{sec:auditory_processing}
Building on auditory awareness, LALMs must also excel in fundamental auditory tasks, such as speech recognition, audio classification, and music analysis, to support advanced real-world applications.  A list of commonly evaluated tasks and their corresponding datasets is provided in Appendix~\ref{sec:task_example} for reference. Initially driven by representation learning models~\citep{wav2vec2, hsu2021hubert, li2022map}, enriched datasets~\citep{pratap2020mls, esc50, maestro}, and existing benchmarks~\citep{yang2021superb, turian2022hear, yuan2023marble}, recent works adapt these resources into instruction-oriented evaluation frameworks tailored for LALMs.

Dynamic-SUPERB~\citep{dynamicsuperb1} initiated this direction, constructing 55 multiple-choice question-answering (QA) tasks spanning speech, audio, and music modalities. Subsequent efforts, such as AIR-Bench~\citep{airbench} and AudioBench~\citep{audiobench}, extend to open-ended QA formats. MuChoMusic~\citep{muchomusic} specifically emphasizes music-related tasks, while Dynamic-SUPERB Phase-2~\citep{dynamicsuperb2} significantly enlarges the benchmark to 180 tasks, forming the largest evaluation suite for LALMs' general processing abilities to date.

Given the task diversity, various evaluation metrics are adopted depending on the task specificity, such as word error rate for speech recognition and BLEU score~\citep{bleu} for translation. There is also an emerging trend that includes LLM-as-a-judge~\citep{llmjudge} for scalable, automatic evaluation of open-ended responses~\citep{dynamicsuperb2, airbench, audiobench}.

Despite achieving promising results in certain areas, these benchmarks demonstrate that current LALMs still fall short of universally robust performance across auditory-processing tasks~\citep{dynamicsuperb2}, highlighting substantial room for improvement toward truly auditory foundation models.

\section{Knowledge and Reasoning}
\label{sec:knowledge_and_reasoning}
Intelligent LALMs should demonstrate extensive knowledge and advanced reasoning to tackle complex real-world tasks. Current evaluations emphasize these abilities through three categories: \textbf{Linguistic Knowledge}, \textbf{World Knowledge Assessment}, and \textbf{Reasoning}. Each category targets distinct but complementary skills, collectively providing a comprehensive evaluation. These assessments reveal key challenges LALMs face in mastering knowledge and reasoning for advanced tasks.

\subsection{Linguistic Knowledge}
\label{sec:linguistic}
Linguistic knowledge refers to understanding and effectively using spoken language. Evaluating LALMs' linguistic proficiency typically use likelihood-based benchmarks where models choose the more linguistically plausible option from paired speech samples. These tests cover lexical knowledge, syntax, and semantic coherence.

Representative works include the ZeroSpeech 2021 benchmark~\citep{zerospeech}, which consists of multiple tracks for evaluating linguistic capabilities. The lexical-level assessment track, sWUGGY, tests models' ability to distinguish between real words and phonotactically similar nonwords, while the syntactic sensitivity evaluation track, sBLIMP, focuses on differentiating grammatical from ungrammatical sentences. CSZS~\citep{cszs} extends syntactic evaluation to multilingual and code-switched scenarios. Narrative and semantic coherence are evaluated by tasks like sStoryCloze and tStoryCloze~\citep{twist}, where models are tasked with selecting semantically appropriate continuations to spoken stories.


\subsection{World Knowledge Assessment}
\label{sec:world_knowledge}
Real-world tasks often demand integrating external knowledge beyond basic auditory understanding. World knowledge assessment evaluates LALMs on two main aspects: (1) auditory expertise like music structure and medical sound diagnosis, and (2) general commonsense and factual knowledge.

Benchmarks that evaluate auditory expertise include MuChoMusic~\citep{muchomusic} and MMAU~\citep{mmau}, which focus on musical understanding, such as melodic structure, harmony, instrument identification, and contextual music interpretation. Additionally, SAGI~\citep{roadmap} assesses medical expertise, such as recognizing illnesses from audio cues like coughing.

Commonsense and factual knowledge evaluations often convert established text benchmarks into spoken form using text-to-speech (TTS). VoxEval~\citep{voxeval} and VoiceBench serve as spoken counterparts to MMLU~\citep{mmlu} and MMLU-Pro~\citep{mmlupro},  testing models across diverse factual domains like social science and humanities. Audiopedia~\citep{audiopedia} uses knowledge graphs from Wikidata~\citep{wikidata} to create audio-based, knowledge-intensive QA tasks that evaluate models' knowledge of well-known entities, such as brands, mentioned in audio.


These benchmarks thoroughly assess LALMs’ knowledge acquisition, revealing challenges such as limited auditory expertise~\citep{muchomusic} and inconsistent performance across domains. Different LALMs excel in different domains, each with their own strengths, but their performance often noticeably declines outside their own specialized areas~\citep{voxeval}. Overall, there remains substantial room to improve LALMs’ auditory expertise and factual knowledge.

\subsection{Reasoning}
\label{sec:reasoning}
Reasoning over auditory inputs falls into two types. Content-based reasoning tests a model’s ability to understand spoken semantic content and answer questions. Acoustic-based reasoning requires utilizing acoustic features like speaker traits and environmental sounds beyond semantics. We provide an overview of these two evaluation paradigms.


\subsubsection{Content-based Reasoning}
Content-based reasoning assesses LALMs' ability to reason over the semantic content of auditory queries. Current benchmarks for this capability typically transform NLP reasoning benchmarks into spoken questions via TTS and require LALMs to provide answers. For instance, VoxEval~\citep{voxeval}, URO-Bench~\citep{urobench}, and ADU-Bench~\citep{adubench} convert NLP datasets like GSM8K~\citep{gsm8k} and MMLU~\citep{mmlu} into speech, evaluating LALMs’ mathematical reasoning based on spoken questions. During synthesis, various speaking styles (e.g., mispronunciation, disfluencies, and accents) may be introduced to test models' robustness~\citep{voxeval}.

These benchmarks reveal gaps in current LALMs' content-based reasoning abilities, even with chain-of-thought~\citep{cot, kojima2022large}. Moreover, model performance varies significantly across speaking styles~\citep{voxeval}, indicating instability in their reasoning.

\subsubsection{Acoustic-based Reasoning}
Acoustic-based reasoning requires LALMs to infer from acoustic cues in auditory input, often involving reasoning across multiple auditory modalities or combining auditory understanding with cognitive skills such as compositional, temporal, logical, and multi-hop reasoning.

\textbf{Cross-auditory Modality Reasoning} demands joint reasoning over multiple auditory modalities, like speech and non-speech sounds. \citet{JASCO} propose an open-ended QA benchmark assessing co-reasoning on speech and environmental sounds, requiring reasoning over cues from distinct auditory sources to infer speakers’ activities. Their findings show that current LALMs frequently neglect non-speech cues, leading to failures.

\textbf{Compositional and Temporal Reasoning} involves comprehending structured acoustic events, their temporal relationships, and attribute binding. Benchmarks like CompA~\citep{compa} evaluate these abilities through specific tasks: CompA-order challenges models to identify correct event sequences or align audio temporal structures with textual descriptions, while CompA-attribute focuses on associating sound events with their sources and attributes. MMAU~\citep{mmau} assesses temporal reasoning via event counting and duration comparison.

\textbf{Logical reasoning} covers structured inference, including deductive and causal reasoning. Deductive reasoning can be tested by Audio Entailment~\citep{audio_entailment}, which evaluates whether a textual hypothesis logically follows from auditory input based on acoustic attributes like sound sources. MMAU~\citep{mmau} examines LALMs' causal reasoning on cause-and-effect relationships of events.

\textbf{Multi-hop reasoning} is the ability to recall and integrate multiple information to answer complex queries, enabling models to connect stored knowledge without explicit reasoning steps~\citep{yang2024large, yang2024largeshortcut, biran-etal-2024-hopping}. SAKURA~\citep{sakura} evaluates LALMs’ multi-hop reasoning by requiring integration of auditory attributes (e.g., speaker gender and emotion) with stored knowledge. Results show that LALMs struggle to combine auditory information with stored knowledge for reasoning, even when both types of information are extracted and known by the models.

\section{Dialogue-oriented Ability}
\label{sec:dialogue}

While foundational skills such as auditory awareness~\Cref{sec:auditory_understanding}, fundamental processing~\Cref{sec:auditory_processing}, language proficiency~\Cref{sec:linguistic}, advanced knowledge~\Cref{sec:world_knowledge}, and reasoning~\Cref{sec:reasoning} are essential for LALMs, natural human-AI interactions additionally require affective and contextual interaction, fluent dialogue management, and precise instruction following. This category targets these integrative skills, focusing on naturalness and controllability, which we group as \textbf{Conversational Ability} and \textbf{Instruction Following}.

\subsection{Conversational Ability}
\label{sec:conversational}

Effective conversational ability in LALMs relies on generating contextually appropriate responses and smoothly managing dialogues in real time. Current evaluations address this via two complementary frameworks: affective and contextual interaction, and full-duplex dialogue management.

\subsubsection{Affective and Contextual Interaction}
Evaluations of affective and contextual interaction typically adopt half-duplex settings, focusing on fully turn-by-turn conversations without speaker overlaps. These benchmarks emphasize LALMs’ ability to respond using both content and non-content cues such as emotional tone, speaking style, and speaker traits. StyleTalk~\citep{styletalk} presents models with a dialogue history and the user’s current speech segment, intentionally leaving the user’s intent underspecified when relying solely on the content. Consequently, models are required to leverage paralinguistic cues to respond appropriately. Subsequent works, such as SD-Eval~\citep{sdeval} and VoxDialogue~\citep{voxdialogue}, broaden the evaluation by incorporating more acoustic and contextual variables, including speaker age, accent, and environmental conditions. These benchmarks combine objective metrics (e.g., ROUGE-L~\citep{rouge}, METEOR~\citep{meteor}), LLM-based judgment~\citep{llmjudge}, and human evaluation for comprehensive assessment.

While these benchmarks rely on static data, \citet{mindthegap} proposes an interactive framework inspired by Chatbot Arena~\citep{chatbotarena}, where real users converse with models on topics of their choice and provide pairwise model preferences, enabling dynamic, user-centered evaluation.

\subsubsection{Full-duplex Dialogue Management}
\label{sec:fullduplex}

Full-duplex evaluation examines LALMs in real-time, dynamic dialogues with complex behaviors like turn-taking~\citep{duncan1972some, gravano2011turn}, backchanneling~\citep{schegloff1982discourse}, and speaker interruptions and overlaps~\citep{interupt, schegloff2000overlapping}. These behaviors are detailed in Appendix~\ref{sec:dynamics}.

Representative works, such as Talking Turns~\citep{talkingturns} and Full-Duplex-Bench~\citep{fullduplexbench}, commonly evaluate four key dimensions:
\begin{itemize}
    \item \textbf{Timing for speaking up or interrupting}: Assesses LALMs’ ability to distinguish meaningful pauses from turn-yielding moments, avoiding undesired interruptions and taking over turns appropriately.

    \item \textbf{Backchanneling}: Evaluates whether LALMs backchannel at proper moments with suitable frequency, reflecting their active listening.

    \item \textbf{Turn taking}: Examines whether LALMs transition smoothly between turns by recognizing boundaries, managing latency, and signaling their intent to maintain or yield the floor.

    \item \textbf{User interruption handling}: Assesses LALMs' handling of interruption, e.g., pausing and smoothly resuming the conversation.
\end{itemize}

Both use automatic evaluation metrics. Talking Turns uses supervised models trained on human dialogues~\citep{godfrey1992switchboard} as a reference, while Full-Duplex-Bench uses metrics like response latency. However, these methods often rely on heuristics, which may be inaccurate in some cases.

Their results show that LALMs struggle with full-duplex management, especially with interruptions~\citep{talkingturns} and seamless turn transitions~\citep{fullduplexbench}, highlighting current limitations in dynamic spoken interaction.

\subsection{Instruction Following}
\label{sec:instruction_following}
Instruction following is the ability to follow user-specified instructions, e.g., requirements for performing particular actions, adhering to constraints, and adjusting response styles. Effective instruction following is essential for model controllability.

LALM instruction-following evaluations typically involve three approaches: (1) adding constraints to existing LALM benchmarks not originally for instruction following, (2) synthesizing LLM instruction-following benchmarks into speech, or (3) creating new dedicated datasets. For instance, Speech-IFeval~\citep{speech_ifeval} introduces constraints into LALM benchmarks such as Dynamic-SUPERB Phase-2~\citep{dynamicsuperb2}; VoiceBench~\citep{voicebench} synthesizes IFEval~\citep{ifeval}, a text-based LLM instruction-following benchmark, into speech; and URO-Bench~\citep{urobench} creates custom evaluation datasets.

Evaluating instruction adherence helps distinguish limitations in following instructions and deficiencies in auditory understanding or knowledge. Common evaluated constraints include length (e.g., a minimum number of words), format (e.g., responses in JSON or all caps), action (e.g., chain-of-thought reasoning~\citep{cot}), style (e.g., responses in a humorous tone), and content (e.g., including a specific word). During evaluation, instruction-following rates, i.e., the frequency with which instructions are correctly followed, are measured with rule-based~\citep{ifeval} or LLM-as-a-judge methods~\citep{llmjudge}.

Benchmark results reveal significant gaps in LALMs compared to their LLM backbones in instruction following~\citep{speech_ifeval}, indicating catastrophic forgetting when adapting LLMs to auditory modalities.


\section{Fairness, Safety, and Trustworthiness}
\label{sec:fairness_safety_trustworthiness}
Despite the advancements of LALMs, their real-world deployment may pose social risks, such as perpetuating biases, generating harmful content, or spreading misinformation, if not properly evaluated and regulated. Therefore, fairness, safety, and trustworthiness must be thoroughly assessed. This section reviews works that quantify these risks to ensure the responsible and ethical use of LALMs.

\subsection{Fairness and Bias}
\label{sec:fairness_bias}
Fairness and bias are key ethical concerns for LALMs, ensuring they do not reinforce societal inequalities, discrimination, stereotypes, or biases. Such issues can be triggered by either the speech content or its non-content acoustic cues. For example, content-triggered bias may arise when LALMs translate occupation-related terms in the speech content into stereotypical gendered terms, independent of acoustic characteristics. In contrast, acoustic-triggered bias may arise when vocal cues lead the model to associate a speaker’s gender with certain occupations.

\citet{lin2024listen} quantifies LALMs' content-triggered gender biases via four tasks: speech-to-text translation, coreference resolution, sentence continuation, and question answering. In each task, gender biases and stereotypes are measured based on the models' responses.

Conversely, Spoken Stereoset~\citep{lin2024spoken} assesses acoustic-triggered bias on speakers' gender and age. The authors sampled sentences from NLP datasets like Stereoset~\citep{stereoset} and BBQ~\citep{parrish-etal-2022-bbq}, which were then rewritten in the first-person perspective with explicit gender or age indicators (e.g., “mother”) removed to ensure bias would be triggered by speaker characteristics rather than content. The modified sentences were synthesized into speech using TTS with voices of different genders and ages. These spoken sentences served as the context, and LALMs were tasked with selecting continuations from options that were stereotypical, anti-stereotypical, or unrelated to the context.


These works highlight LALMs' social biases, which may be inherited from their training data or LLM backbones. Additionally, since social biases are multifaceted, current benchmarks cannot include all possible societal factors, emphasizing the need for further research into both model development and benchmarks to enhance fairness.

\subsection{Safety}
\label{sec:safety}
Unlike fairness and bias, which expose societal prejudices in LALMs, safety concerns focus on preventing harmful or unsafe outputs that may negatively impact individuals or society, including user discomfort or illegal activities. Current studies typically use NLP datasets with malicious queries and convert them into speech via TTS. For example, VoiceBench~\citep{voicebench} and~\citet{multiaudiojail} synthesize datasets like AdvBench~\citep{advbench} into spoken queries, evaluating LALMs on their ability to reject them.

During evaluation, jailbreaking techniques may be employed to test models’ resistance to adversarial inputs. These include modifying speech content by inserting fictional scenarios~\citep{shen2024voice} and applying auditory manipulations such as silence~\citep{yang-etal-2025-audio}, noise~\citep{yang-etal-2025-audio, xiao2025tune}, accents~\citep{multiaudiojail, xiao2025tune}, and audio edits~\citep{xiao2025tune, gupta2025bad}. Ideally, LALMs should remain robust to adversarially modified inputs and consistently reject malicious requests.

However, evaluations show that LALMs often accept malicious spoken inputs even when they can refuse similar textual ones~\citep{voicebench}. Moreover, LALMs show considerable safety degradation compared to their LLM backbones~\citep{yang-etal-2025-audio}. Several jailbreaking methods can easily bypass these models~\citep{multiaudiojail, xiao2025tune}, highlighting the need for better multimodal safety alignment.

\subsection{Hallucination}
\label{sec:hallucination}

Hallucination occurs when a model generates nonfactual or unsupported outputs, reducing reliability and misleading users. In LALMs, hallucinations can originate from both auditory and textual modalities. While textual hallucinations can be assessed with NLP benchmarks~\citep{li2023halueval, chen-etal-2024-diahalu, bang2025hallulens}, we focus on auditory-induced hallucinations.

\citet{kuan24_interspeech} explores LALMs' object hallucination, where the models falsely identify objects or events absent from the auditory input. They evaluate this via two tasks: a discriminative task where LALMs determine whether a specified object exists in the audio, and a generative task where LALMs generate captions describing the audio. These captions are then evaluated for accuracy in reflecting the actual content of the audio. Despite generating accurate captions, LALMs struggle with object identification in the discriminative task, revealing challenges in object hallucination for question-answering tasks.

\citet{cmm} further analyzes object hallucination using the CMM benchmark, showing that overrepresented objects or events in the training data can lead LALMs to incorrectly predict their presence, even when they are absent. Additionally, the frequent co-occurrence of objects and events during training exacerbates these hallucinations. 

These works highlight hallucination challenges in LALMs and call for improved training, modeling, and data handling to enhance trustworthiness.



\section{Challenges and Future Directions}
\label{sec:challenges}


\subsection{Data Leakage and Contamination}
\label{sec:data_leakage}
Creating and curating high-quality auditory data is far more difficult than for text. Consequently, many LALM benchmarks rely on existing auditory corpora~\citep{librispeech, audiocaps, audioset} rather than collecting new data. This raises concerns about data leakage, since models may have seen these datasets during training~\citep{deng-etal-2024-investigating, zhou2023don, jacovi2023stop}, undermining evaluation reliability. The risk grows when large-scale web-crawled data~\citep{whisper, he2024emilia} are used for training without rigorous filtering.

Thus, alongside creating or collecting custom data, developing methods to detect and mitigate contamination~\citep{golchin2024time, samuel2025towards} will be a crucial direction for more reliable LALM evaluations.

\subsection{Inclusive Evaluation Across Linguistic, Cultural, and Communication Diversity}

While current benchmarks cover major languages like English and Mandarin~\citep{dynamicsuperb2, urobench}, many overlook crucial aspects such as low-resource languages~\citep{magueresse2020low} and code-switching~\citep{dougruoz2021survey, sitaram2019survey}. Although these have been explored in traditional speech technologies~\citep{khare2021low, bhogale24_interspeech, liu2024enhancing, csasrst}, they remain underexamined in LALMs. This limited coverage fails to capture the full linguistic diversity of human communication, as different languages possess unique characteristics~\citep{evans2009myth, bickel2014linguistic}.

Cultural factors, shaped by historical and social contexts, influence dimensions like moral norms~\citep{graham2016cultural, saucier2018culture} and are essential for evaluation. As LALMs extend to diverse cultures~\citep{yang2024building, wang2025advancing}, evaluation frameworks must also expand.

Along with language and culture, communication patterns also matter. While some work covers speech variations like accents, underrepresented groups such as people with speech disorders (e.g., dysarthria~\citep{kent1999acoustic, kim2008dysarthric}) are often overlooked, as current LALMs have limited familiarity with their unique speech patterns, which affects fair and accurate understanding.

To develop fair and broadly applicable LALMs, future evaluations should carefully consider linguistic, cultural, and communicative diversity.

\subsection{Safety Evaluation Unique to Auditory Modalities}

Current LALM safety evaluations~\Cref{sec:safety} mainly target harmful content in model outputs, often overlooking risks inherent to auditory modalities. Auditory cues such as tone, emotion, and voice quality can also influence user experience and raise concerns if uncontrolled. For instance, even harmless content can discomfort users if spoken harshly or sarcastically, and the presence of annoying noises can also cause irritation. Thus, safety should cover auditory comfort, not just content harmlessness.

Most benchmarks focus on content toxicity but seldom assess auditory-specific safety. Addressing these issues is vital for applications like voice assistants~\citep{building_trust, mari2024empathic}, where vocal manner greatly affects user trust and comfort. Future work should jointly consider vocal tone, noise, and other paralinguistic factors to ensure safe, user-friendly interactions.

\subsection{Unified Evaluation of Harmlessness and Helpfulness}

Harmlessness and helpfulness in LALMs refer to safety and fairness, and the ability to assist users, respectively. Ideally, these two properties should be enhanced together; however, in practice, they often conflict~\citep{bai2022training}. For example, a model that always refuses to answer is safe but unhelpful, as it fails to assist users. A recent study~\citep{lin2025preliminary} shows that post-training aimed at enhancing harmlessness can reduce helpfulness, causing models to reject queries even when no safety or privacy issues exist. This tension highlights the need for a unified evaluation framework that considers both aspects simultaneously.

Existing harmlessness benchmarks~\cref{sec:fairness_safety_trustworthiness} rarely include helpfulness, limiting understanding of their trade-offs and offering limited guidance for balancing them effectively. Thus, developing a joint evaluation framework is a key future direction.

\subsection{Personalization Evaluation}


Personalization enables models to adapt to individual users by incorporating private information like users' voices and preferences, supporting applications such as personalized voice assistants.

While traditional speech technologies have explored personalization~\citep{ASRpersonalization, joseph2024speaker}, it remains underdeveloped for LALMs. Unlike recent progress in LLM personalization~\citep{tan2024personalized, tan2025personabench, zhang2024personalization}, LALM personalization is more complex due to the auditory dimension: LALMs must adapt to user-specific knowledge, as text LLMs do, but also become familiar with users’ voice characteristics and speaking habits, and adjust their own speaking style to match user preferences. Such complexity necessitates the development of specialized evaluations to fully assess LALM personalization, making it a valuable area for future investigation.

\section{Conclusion}

Holistic evaluation of LALMs is as crucial as modeling and training in advancing the field. This survey reviews existing evaluation frameworks and proposes a taxonomy categorizing current progress into four important research areas, reflecting the diverse expectations of LALM capabilities. We present a thorough overview of the literature, highlighting challenges and future directions, such as data contamination, inclusivity, auditory-specific safety, and personalization. We hope this survey provides clear guidelines for researchers and stimulates further advancements in LALM evaluation.

\section*{Limitations}
\label{sec:limitations}
We acknowledge a few limitations in this paper. First, the scope of our taxonomy is based on existing evaluation frameworks and benchmarks, meaning it does not cover all possible real-world auditory tasks. The auditory modalities are inherently complex, with a wide range of tasks and applications that cannot be exhaustively covered. As LALMs continue to evolve, new capabilities and applications will emerge, leading to growing expectations for these models. Consequently, the evaluation landscape will likely expand and shift, requiring our taxonomy to be updated and adapted to include these new tasks and applications. We will continue to follow the advancements in this field and adjust our taxonomy accordingly to reflect these developments.

Second, this survey primarily focuses on current benchmarks used to evaluate LALMs' performance across various aspects. As a result, it does not put much emphasis on more basic or traditional evaluation methods, such as subjective assessments of speech generation quality (e.g., Mean Opinion Score), which are commonly used to evaluate model-generated audio. While these methods are valuable in certain applications, they fall outside the scope of this paper, which aims to provide a comprehensive overview of more advanced and specialized benchmarks.

\section*{Acknowledgments}
We thank the reviewers for their valuable feedback, which have helped us improve the paper. To enhance the comprehensiveness of our survey, we have added further discussion in the appendix. This work was supported by the Ministry of Education (MOE) of Taiwan under the project Taiwan Centers of Excellence in Artificial Intelligence, through the NTU Artificial Intelligence Center of Research Excellence (NTU AI-CoRE)

\bibliography{main}

\appendix

\section{Detailed Categorization of the Surveyed Papers}
\label{sec:entiretaxonomy}
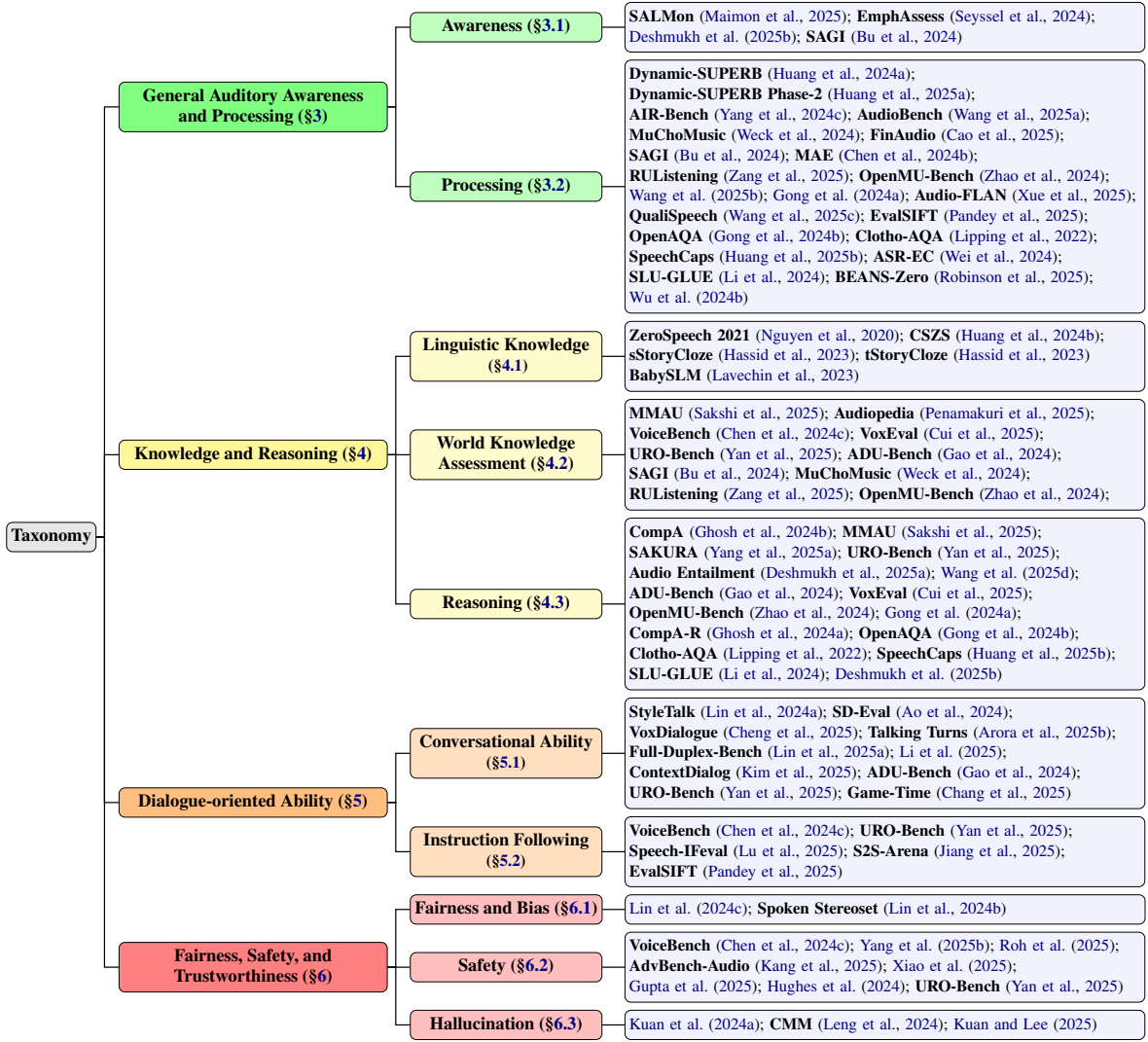
\begin{figure*}[t!]
\centering
\resizebox{0.97\textwidth}{!}{%
\begin{forest}
 for tree={
    grow'=0,
    parent anchor=east, child anchor=west,
    edge={draw, line width=0.8pt},
    forked edges,
    align=center, anchor=west,
    l sep=15pt, s sep=5pt, inner sep=3pt,
    minimum width=5cm, draw, fill=blue!5, rounded corners,
    where n children=3{
      calign=child edge,
      calign child=2,
    }{},
  }
  [{\large\textbf{Taxonomy}}, minimum width=2.1cm, fill=gray!20
    [\large\textbf{General Auditory Awareness}\\\large\textbf{and Processing~\Cref{sec:understanding_and_processing}}, fill=green!50, minimum width=6.3cm
      [\large\textbf{Awareness~\Cref{sec:auditory_understanding}}, fill=green!25, minimum width=4.5cm
        [\textbf{SALMon}~\citep{salmonbench}; \textbf{EmphAssess}~\citep{seyssel2024emphassess};\\
        \citet{deshmukh2025adiff}; \textbf{SAGI}~\citep{roadmap}
        ,align=left, text width=12cm, minimum width=12cm]
      ]
      [\large\textbf{Processing~\Cref{sec:auditory_processing}}, fill=green!25, minimum width=4.5cm
        [\textbf{Dynamic-SUPERB}~\citep{dynamicsuperb1};\\ \textbf{Dynamic-SUPERB Phase-2}~\citep{dynamicsuperb2};\\
          \textbf{AIR-Bench}~\citep{airbench}; \textbf{AudioBench}~\citep{audiobench};\\
          \textbf{MuChoMusic}~\citep{muchomusic}; \textbf{FinAudio}~\citep{cao2025finaudio};\\
          \textbf{SAGI}~\citep{roadmap}; \textbf{MAE}~\citep{mae};\\
          \textbf{RUListening}~\citep{rulistening}; \textbf{OpenMU-Bench}~\citep{openmubench};\\
          \citet{wang2025advancing}; \citet{gong2024av}; \textbf{Audio-FLAN}~\citep{audioflan};\\
          \textbf{QualiSpeech}~\citep{qualispeech}; \textbf{EvalSIFT}~\citep{evalsift};\\
          \textbf{OpenAQA}~\citep{openaqa}; \textbf{Clotho-AQA}~\citep{clothoaqa};\\
          \textbf{SpeechCaps}~\citep{speechcaps}; \textbf{ASR-EC}~\citep{asrec};\\
          \textbf{SLU-GLUE}~\citep{slu-glue}; \textbf{BEANS-Zero}~\citep{beans-zero};\\
          \citet{wu2024just}
         ,align=left, text width=12cm, minimum width=12cm]
      ]
    ]
    [\large\textbf{Knowledge and Reasoning~\Cref{sec:knowledge_and_reasoning}}, fill=yellow!50, minimum width=6.3cm
      [\large\textbf{Linguistic Knowledge}\\\textbf{
      \Cref{sec:linguistic}}, fill=yellow!25, minimum width=4.5cm
        [{\textbf{ZeroSpeech 2021}~\citep{zerospeech}; \textbf{CSZS}~\citep{cszs}};\\\textbf{sStoryCloze}~\citep{twist}; \textbf{tStoryCloze}~\citep{twist}\\
        \textbf{BabySLM}~\citep{babyslm}
         ,align=left, text width=12cm, minimum width=12cm]
      ]
      [\large\textbf{World Knowledge}\\\large\textbf{Assessment~\Cref{sec:world_knowledge}}, fill=yellow!25, minimum width=4.5cm
        [\textbf{MMAU}~\citep{mmau}; \textbf{Audiopedia}~\citep{audiopedia};\\
          \textbf{VoiceBench}~\citep{voicebench}; \textbf{VoxEval}~\citep{voxeval};\\\textbf{URO-Bench}~\citep{urobench}; \textbf{ADU-Bench}~\citep{adubench};\\
          \textbf{SAGI}~\citep{roadmap}; \textbf{MuChoMusic}~\citep{muchomusic}; \\\textbf{RUListening}~\citep{rulistening}; \textbf{OpenMU-Bench}~\citep{openmubench};,
         align=left, text width=12cm, minimum width=12cm]
      ]
      [\large\textbf{Reasoning~\Cref{sec:reasoning}}, fill=yellow!25, minimum width=4.5cm
        [\textbf{CompA}~\citep{compa}; \textbf{MMAU}~\citep{mmau};\\\textbf{SAKURA}~\citep{sakura}; \textbf{URO-Bench}~\citep{urobench};\\\textbf{Audio Entailment}~\citep{audio_entailment}; \citet{JASCO};\\\textbf{ADU-Bench}~\citep{adubench}; \textbf{VoxEval}~\citep{voxeval};\\
        \textbf{OpenMU-Bench}~\citep{openmubench}; \citet{gong2024av}; \\
        \textbf{CompA-R}~\citep{gama_compaR}; \textbf{OpenAQA}~\citep{openaqa};\\
        \textbf{Clotho-AQA}~\citep{clothoaqa}; \textbf{SpeechCaps}~\citep{speechcaps};\\
        \textbf{SLU-GLUE}~\citep{slu-glue}; \citet{deshmukh2025adiff}
         ,align=left, text width=12cm, minimum width=12cm]
      ]
    ]
    [\large\textbf{Dialogue-oriented Ability~\Cref{sec:dialogue}}, fill=orange!50, minimum width=6.3cm
      [\large\textbf{Conversational Ability}\\\textbf{\Cref{sec:conversational}}, fill=orange!25, minimum width=4.5cm
        [\textbf{StyleTalk}~\citep{styletalk}; \textbf{SD-Eval}~\citep{sdeval};\\
          \textbf{VoxDialogue}~\citep{voxdialogue}; \textbf{Talking Turns}~\citep{talkingturns};\\
          \textbf{Full-Duplex-Bench}~\citep{fullduplexbench}; \citet{mindthegap};\\
          \textbf{ContextDialog}~\citep{contextdialog}; \textbf{ADU-Bench}~\citep{adubench}; \\\textbf{URO-Bench}~\citep{urobench}; \textbf{Game-Time}~\citep{gametime}
          ,align=left, text width=12cm, minimum width=12cm]
      ]
      [\large\textbf{Instruction Following}\\\textbf{\Cref{sec:instruction_following}}, fill=orange!25, minimum width=4.5cm
        [\textbf{VoiceBench}~\citep{voicebench}; \textbf{URO-Bench}~\citep{urobench};\\\textbf{Speech-IFeval}~\citep{speech_ifeval}; \textbf{S2S-Arena}~\citep{s2sarena};\\
        \textbf{EvalSIFT}~\citep{evalsift},
         align=left, text width=12cm, minimum width=12cm]
      ]
    ]
    [\large\textbf{Fairness, Safety, and} \\\large\textbf{Trustworthiness~\Cref{sec:fairness_safety_trustworthiness}}, fill=red!50, minimum width=6.3cm
      [\large\textbf{Fairness and Bias~\Cref{sec:fairness_bias}}, fill=red!25, minimum width=4.5cm
        [{\citet{lin2024listen}; \textbf{Spoken Stereoset}~\citep{lin2024spoken}},
         align=left, text width=12cm, minimum width=12cm]
      ]
      [\large\textbf{Safety~\Cref{sec:safety}}, fill=red!25, minimum width=4.5cm
        [\textbf{VoiceBench}~\citep{voicebench}; \citet{yang-etal-2025-audio}; \citet{multiaudiojail};\\
        \textbf{AdvBench-Audio}~\citep{advbench_audio}; \citet{xiao2025tune};\\\citet{gupta2025bad}; \citet{hughes2024best}; \textbf{URO-Bench}~\citep{urobench},
         align=left, text width=12cm, minimum width=12cm]
      ]
      [\large\textbf{Hallucination~\Cref{sec:hallucination}}, fill=red!25, minimum width=4.5cm
        [{\citet{kuan24_interspeech}; \textbf{CMM}~\citep{cmm}; \citet{match}},
         align=left, text width=12cm, minimum width=12cm]
      ]
    ]
  ]
\end{forest}
}
\caption{The complete categorization of the surveyed papers based on the proposed taxonomy.}
\label{fig:entire_taxonomy}
\end{figure*}
The complete categorization of the surveyed papers, based on the proposed taxonomy~\cref{sec:taxonomy}, is presented in Figure~\ref{fig:entire_taxonomy}. Please note that widely used corpora for fundamental auditory processing tasks, such as speech recognition and audio captioning, are excluded from this categorization due to the extremely large number of such resources. Including them would make the figure overly detailed and cumbersome. For reference, we provide examples of these fundamental tasks and their corresponding resources in Appendix~\ref{sec:task_example}.

From Figure~\ref{fig:entire_taxonomy}, it is evident that the current focus of LALM evaluations predominantly centers on auditory processing tasks~\cref{sec:auditory_processing}, underscoring their importance to the community. While these tasks are valuable, they should not be seen as the sole consideration when evaluating models for real-world applications. A more diverse and comprehensive evaluation scope is crucial to ensure a fuller understanding of their potential and shortcomings.

\section{Brief Summary for Benchmarks Discussed in the Main Text}

In this section, we summarize the key features and evaluation metrics of the benchmarks presented in Figure~\ref{fig:taxonomy}, aiming to guide researchers in selecting benchmarks suitable for their own use cases. The details are provided in Table~\ref{tab:benchmarks_part1} and Table~\ref{tab:benchmarks_part2}.

\begin{table}[t] \scriptsize
\setlength\tabcolsep{0.5pt} 
\renewcommand{\arraystretch}{0.05}

\centering

\resizebox{0.95\linewidth}{!}{
\begin{tabular}{cc}
  \toprule
  \textbf{Auditory Tasks} & \textbf{Common Datasets} \\

  \midrule
  \rowcolor{gray!30} \multicolumn{2}{c}{\textbf{Audio Tasks}}\\
  \midrule

   Audio Captioning  & \makecell[c]{
   AudioCaps~\cite{audiocaps}\\
   Clotho~\cite{drossos2020clotho}
   } \\
    \midrule
    
   Audio Classification & \makecell[c]{
   ESC-50~\cite{ESC-50}\\
   AudioSet~\cite{audioset}\\
   } \\

   \midrule
   Vocal Sound Classification & VocalSound~\citep{vocalsound}\\

  \midrule
  \rowcolor{gray!30} \multicolumn{2}{c}{\textbf{Speech Tasks}}\\
  \midrule

  Automatic Speech Recognition  & \makecell[c]{
  LibriSpeech~\cite{panayotov2015librispeech}\\
  AISHELL-1~\cite{bu2017aishell}\\
  Common Voice~\cite{commonvoice}\\
  } \\
   \midrule
   Speaker Identification  & \makecell[c]{
   VoxCeleb2~\cite{chung2018voxceleb2}
   \\
   CN-Celeb~\cite{CN-Celeb}
   } \\

   \midrule
   Text-to-Speech  & \makecell[c]{
   LJSpeech~\cite{ljspeech17}\\
   VCTK~\cite{CSTR-VCTK}\\
   LibriTTS~\cite{zen2019libritts}
   } \\

   \midrule
   Speech Emotion Recognition  & \makecell[c]{
   IEMOCAP~\cite{IEMOCAP2008}\\
   CREMA-D~\cite{crema-d}
   } \\

   \midrule
   Language Identification  & \makecell[c]{
   VoxLingua107~\cite{valk2021voxlingua107}\\
   FLEURS~\cite{conneau2023fleurs}
   } \\

   \midrule
   Speech Translation  & \makecell[c]{
   CoVoST 2~\cite{wang2021covost}\\
   MuST-C ~\cite{di-gangi-etal-2019-must}
   } \\

   \midrule
   Speech Diarization  & \makecell[c]{
    LibriMix~\cite{cosentino2020librimix}
   } \\
    \midrule
   Keyword Spotting  & \makecell[c]{
    Speech Command~\cite{speechcommand}
   } \\


  \midrule
  \rowcolor{gray!30} \multicolumn{2}{c}{\textbf{Music Tasks}}\\
  \midrule
  
  \makecell[c]{Music Captioning\\Text-to-Music}  & \makecell[c]{
    MusicCaps~\cite{MusicLM}\\
    Song Describer Dataset~\cite{manco2023song}\\
    MidiCaps~\cite{melechovsky2024midicaps} \\
   }\\
   \midrule

   Music Transcription  & \makecell[c]{
    MAESTRO~\cite{maestro} 
   }\\
   \midrule

  Instrument Classification & NSynth~\citep{nsynth}\\
  \midrule

  Genre Classification & \makecell[c]{
    FMA~\citep{fma}\\
    GTZAN~\citep{gtzan}
   }\\

  \bottomrule

\end{tabular}
}
\caption{Commonly used datasets for various auditory tasks. This overview covers key tasks in audio, speech, and music processing and the datasets that are widely adopted in academic and industrial research.}
\label{tab:task_example}

\end{table}
\begin{table}[!ht]
\setlength\tabcolsep{3pt}
\renewcommand{\arraystretch}{0.3}
\centering
\resizebox{0.9\linewidth}{!}{
\begin{tabular}{cc}
\toprule


\textbf{Dataset} & \makecell[c]{\textbf{\# of Benchmarks Using}\\ \textbf{the Dataset}} \\

\midrule


\textbf{AudioCaps} & \textbf{6} \\

\midrule


\textbf{Clotho} & \textbf{5} \\

\midrule


ESC-50 & 3 \\

\midrule


\textbf{AudioSet} & \textbf{8} \\

\midrule

VocalSound & 2 \\

\midrule

\textbf{LibriSpeech} & \textbf{7} \\

\midrule

\textbf{Common Voice} & \textbf{8} \\

\midrule

VoxCeleb (1\&2) & 3 \\

\midrule

LJSpeech & 3 \\

\midrule

VCTK & 4 \\

\midrule

LibriTTS & 2 \\

\midrule

IEMOCAP & 4 \\

\midrule

CREMA-D & 2 \\

\midrule

VoxLingua107 & 1\\

\midrule

CoVoST 2 & 2 \\

\midrule

LibriMix & 1 \\

\midrule

Speech Command & 2 \\

\midrule

MusicCaps & 4 \\

\midrule

Song Describer Dataset & 3 \\

\midrule

MAESTRO & 1 \\

\midrule

NSynth & 2 \\

\midrule

FMA & 2 \\

\bottomrule
\end{tabular}
}
\caption{Number of benchmarks in Figure~\ref{fig:taxonomy} that use the datasets in Table~\ref{tab:task_example}. Datasets used by more than five benchmarks are highlighted in bold, while those not used by any benchmark are omitted.}
\label{tab:data_overlap}
\end{table}
\begin{table}[t]
\centering
\scriptsize
\renewcommand{\arraystretch}{0.9}
\setlength\tabcolsep{5pt}
\begin{tabularx}{0.99\linewidth}{X c c}
\toprule
\textbf{Benchmark} & \textbf{Real Data} & \textbf{Synthetic Data} \\
\midrule
\makecell[tl]{SALMon \\ \citep{salmonbench}} & \cmark & \cmark \\
\midrule
\makecell[tl]{EmphAssess \\ \citep{seyssel2024emphassess}} &  & \cmark \\
\midrule
\makecell[tl]{Dynamic-SUPERB \\ \citep{dynamicsuperb1}} & \cmark & \cmark \\
\midrule
\makecell[tl]{Dynamic-SUPERB Phase-2 \\ \citep{dynamicsuperb2}} & \cmark & \cmark \\
\midrule
\makecell[tl]{AIR-Bench \\ \citep{airbench}} & \cmark & \cmark \\
\midrule
\makecell[tl]{AudioBench \\ \citep{audiobench}} & \cmark & \cmark \\
\midrule
\makecell[tl]{MuChoMusic \\ \citep{muchomusic}} & \cmark &  \\
\midrule
\makecell[tl]{ZeroSpeech 2021 \\ \citep{zerospeech}} & \cmark & \cmark \\
\midrule
\makecell[tl]{CSZS \\ \citep{cszs}} &  & \cmark \\
\midrule
\makecell[tl]{sStoryCloze \& tStoryCloze \\ \citep{storycloze}} &  & \cmark \\
\midrule
\makecell[tl]{MMAU \\ \citep{mmau}} & \cmark & \cmark \\
\midrule
\makecell[tl]{Audiopedia \\ \citep{audiopedia}} &  & \cmark \\
\midrule
\makecell[tl]{VoiceBench \\ \citep{voicebench}} & \cmark & \cmark \\
\midrule
\makecell[tl]{VoxEval \\ \citep{voxeval}} &  & \cmark \\
\midrule
\makecell[tl]{CompA \\ \citep{compa}} & \cmark & \cmark \\
\midrule
\makecell[tl]{SAKURA \\ \citep{sakura}} & \cmark &  \\
\midrule
\makecell[tl]{URO-Bench \\ \citep{urobench}} & \cmark & \cmark \\
\midrule
\makecell[tl]{Audio Entailment \\ \citep{audio_entailment}} & \cmark &  \\
\midrule
\citet{JASCO} & \cmark & \cmark \\
\midrule
\makecell[tl]{StyleTalk \\ \citep{styletalk}} &
 & \cmark \\
\midrule
\makecell[tl]{SD-Eval \\ \citep{sdeval}} &
\cmark & \cmark \\
\midrule
\makecell[tl]{VoxDialogue \\ \citep{voxdialogue}} &
 &
\cmark \\
\midrule
\makecell[tl]{Talking Turns \\ \citep{talkingturns}} &
\cmark & \\
\midrule
\makecell[tl]{Full-Duplex-Bench \\ \citep{fullduplexbench}} &
\cmark & \cmark \\
\midrule
\makecell[tl]{Speech-IFEval \\ \citep{speech_ifeval}} & \cmark & \\
\midrule
\citet{lin2024listen} &
\cmark & \cmark \\
\midrule
\makecell[tl]{Spoken Stereoset \\ \citep{lin2024spoken}} & & \cmark \\
\midrule
\citet{yang-etal-2025-audio} & & \cmark \\
\midrule
\citet{multiaudiojail} & & \cmark \\
\midrule
\citet{kuan24_interspeech} &
\cmark & \\
\midrule
\makecell[tl]{CMM \\ \citep{cmm}} &
\cmark & \\
\bottomrule
\end{tabularx}
\caption{Statistics of the utilization of real/synthetic auditory data in benchmarks in Figure~\ref{fig:taxonomy}.}
\label{tab:synthetic}
\end{table}

\section{Examples of General Auditory Processing Tasks and Resources}
\label{sec:task_example}


Table~\ref{tab:task_example} lists representative auditory processing tasks and their associated resources. As foundational components of auditory processing, these tasks are well-suited for adaptation in LALM evaluation, as discussed in~\Cref{sec:auditory_processing}.

\section{Overlap between Common Corpora and Existing Benchmarks}
\label{sec:data_overlap}

In Table~\ref{tab:task_example}, we list several corpora that are commonly used in the community. However, as discussed in~\cref{sec:data_leakage}, evaluation benchmarks may face risks of data leakage and contamination if they heavily rely on these existing resources. To provide a quantitative view of this issue, we report detailed statistics on the number of benchmarks discussed in the main text (Figure~\ref{fig:taxonomy}) that make use of the datasets in Table~\ref{tab:task_example}.

From Table~\ref{tab:data_overlap}, we observe that certain corpora are used with particularly high frequency, which reinforces concerns about data contamination, especially when web-scraped data are incorporated into training without rigorous filtering. We emphasize that this issue should be taken seriously.

\section{Utilization of Synthetic Auditory Data in Benchmarks in the Main Text}

Table~\ref{tab:synthetic} summarizes whether the benchmarks in Figure~\ref{fig:taxonomy} use real or synthetic auditory data. Synthetic audio is mainly employed to (1) generate data difficult to obtain in real settings, such as controlled stress or specific sound effects, and (2) verbalize task instructions or dialogues consistently.

\section{Dynamics in Full-Duplex Dialogues}
\label{sec:dynamics}

In this section, we briefly introduce the dynamics discussed in~\cref{sec:fullduplex}. \textbf{Turn-taking}~\citep{sacks1974simplest} is a fundamental aspect of conversational organization, where speakers alternate turns to speak, ensuring only one person talks at a time. This process is complex, involving various behaviors that help facilitate smooth transitions between speakers. For example, speakers often signal the end of their turn through clear cues, allowing the listener to recognize when they are yielding the floor~\citep{duncan1972some, duncan2015face}. Furthermore, turn-taking conventions may be shaped by cultural factors~\citep{sidnell2007comparative}, which influence how and when speakers take their turns due to linguistic and social differences. Understanding and modeling these behaviors are essential steps toward achieving natural and effective communication in both human-human and human-AI interactions.

\textbf{Backchanneling} involves the listener's use of phatic expressions that signal active listening and attentiveness to the speaker~\citep{fujie2005back}. These verbal cues, such as ``yeah,'' ``I see,'' or ``uh-huh,'' along with non-verbal cues like nodding, serve as feedback, showing sympathy, agreement, or understanding. By offering such responses, listeners help maintain the flow of conversation without interrupting the speaker. This behavior not only fosters a sense of connection but also enhances the speaker's feeling of being heard and understood, contributing to a more interactive and supportive dialogue. As such, backchanneling plays a crucial role in sustaining conversation dynamics and promoting positive communicative exchanges.

\textbf{Speaker overlap} refers to the simultaneous speech of multiple speakers, while \textbf{speaker interruption} occurs when one speaker interjects during another's turn, which breaks the turn-taking principles~\citep{interupt}. These phenomena are complex: they can be competitive, reflecting hostility or dominance~\citep{west1979against, orcutt1985deviance}, or they can be neutral or supportive, helping to maintain and coordinate the flow of dialogue~\citep{goldberg1990interrupting, jefferson1986notes, gervits2018towards}. Despite their varying forms, both overlap and interruption are natural components of human conversation.

\section{Input/Output Modalities of the Surveyed Works}

Our proposed taxonomy~\Cref{sec:taxonomy} is organized by the evaluation objectives of the surveyed works rather than by the modalities they cover. Nevertheless, modality information is essential for researchers seeking benchmarks suited to models specialized in particular modalities. Thus, we provide the input/output modality details in Tables~\ref{tab:modality_part1},~\ref{tab:modality_part2},~\ref{tab:modality_part3}, and~\ref{tab:modality_part4}, corresponding to the categories of General Auditory Awareness and Processing~\Cref{sec:understanding_and_processing}, Knowledge and Reasoning~\Cref{sec:knowledge_and_reasoning}, Dialogue-oriented Ability~\Cref{sec:dialogue}, and Fairness, Safety, and Trustworthiness~\Cref{sec:fairness_safety_trustworthiness}, respectively. These tables are compiled based on the original papers of the surveyed works.

Please note that due to unique evaluation designs, some benchmarks do not produce explicit ``outputs'' but instead rely on input likelihood comparisons or similarity measures with specific instances. This absence of outputs is clearly indicated in the tables.


\section{Comparison between End-to-end LALMs and Cascaded Systems}

\begin{table*}[h]
\centering
\begin{threeparttable}
\resizebox{0.99\textwidth}{!}{
\begin{tabular}{c c}
\toprule
\textbf{Type} & \textbf{Benchmark List} \\
\midrule
\multirow{8}{*}{\makecell[c]{Cascaded Systems\\ Outperform}} & Dynamic-SUPERB~\citep{dynamicsuperb1} \\
 & MMAU~\citep{mmau} \\
 & VoiceBench~\citep{voicebench} \\
 & VoxEval~\citep{voxeval} \\
 & SAKURA~\citep{sakura} \\
 & SD-Eval~\citep{sdeval} \\
 & Speech-IFEval~\citep{speech_ifeval} \\
 & \citet{yang-etal-2025-audio} \\

\midrule
\multirow{4}{*}{\makecell[c]{E2E LALMs\\ Outperform}} & SALMon~\citep{salmonbench} \\
 & Dynamic-SUPERB Phase-2~\citep{dynamicsuperb2}\tnote{1} \\
 & AIR-Bench~\citep{airbench}\tnote{2} \\
 & StyleTalk~\citep{styletalk}\tnote{3} \\

\bottomrule
\end{tabular}
}
\begin{tablenotes}\footnotesize
\parbox{0.95\textwidth}{

\item[1] Overall, the top-performing E2E LALMs outperform cascaded systems, except for a small number of tasks where their performance is slightly inferior.

\item[2] The top-performing E2E LALMs significantly outperform cascaded systems on the foundation benchmark, with only a slight drop in performance on the chat benchmark.

\item[3] The top-performing E2E LALMs significantly outperform cascaded systems across various metrics, except for one where they slightly underperform. 
}
\end{tablenotes}
\end{threeparttable}
\caption{Summary of comparisons between end-to-end (E2E) LALMs and cascaded systems, limited to benchmarks in Figure~\ref{fig:taxonomy} whose original papers explicitly conducted and reported such comparisons.}
\label{tab:cascaded}
\end{table*}

Table~\ref{tab:cascaded} presents a comparison between end-to-end (E2E) LALMs and cascaded systems that couple LLMs with modules such as speech recognition, evaluated across the benchmarks in Figure~\ref{fig:taxonomy}. Cascaded systems generally perform better on benchmarks emphasizing content- or semantics-based tasks, which require only limited integration of non-semantic auditory cues. These tasks align well with the strengths of cascaded pipelines. For instance, VoiceBench~\citep{voicebench} and VoxEval~\citep{voxeval} primarily assess world knowledge and reasoning abilities adapted from text-based datasets (e.g., MMLU~\citep{mmlu}), where textual representations alone are sufficient, thus giving cascaded approaches a clear advantage.

Conversely, certain benchmarks highlight the advantages of E2E LALMs. For example, SALMon~\citep{salmonbench} requires fine-grained auditory perception that cascaded systems struggle to replicate. This suggests the importance of future benchmarks that place greater emphasis on nuanced auditory understanding and reasoning, where LALMs hold stronger potential to excel.

\section{Common Metrics for LALMs}

As previously noted, this paper primarily focuses on benchmarks for evaluating LALMs across diverse aspects, rather than on basic or traditional evaluation metrics. In this section, we briefly introduce some of the most commonly used metrics.

To assess the generation quality and naturalness of auditory outputs, the most widely adopted measure is the Mean Opinion Score (MOS), which relies on human annotators’ judgments to provide subjective quality assessments. While effective, MOS collection can be costly and time-consuming. To mitigate this, proxy models such as UTMOS~\citep{utmos} have been proposed to automatically predict MOS scores, offering a more efficient alternative to direct human evaluation.

For evaluating the content of generated auditory outputs, several additional metrics are commonly used. In cases where models can produce both speech and text, Character Error Rate (CER) and Word Error Rate (WER) measure the consistency between generated textual outputs and transcriptions of generated speech (via an ASR system), thereby quantifying alignment across modalities.

Beyond surface-level alignment, other metrics assess the semantic quality of responses. Widely used measures include ROUGE~\citep{rouge} and BERTScore~\citep{zhangbertscore}, which evaluate semantic overlap between model outputs and references. More recently, LLM-as-a-judge~\citep{llmjudge} has been increasingly adopted to provide flexible, criterion-driven evaluations tailored to researchers’ specific needs.

\section{Information on AI Assistance}
We acknowledge the assistance of GPT-4.1-mini in refining the paper and improving its clarity.

\begin{table*}[t]
\centering
\scriptsize
\renewcommand{\arraystretch}{0.9}
\setlength\tabcolsep{4pt}
\begin{tabularx}{0.9\textwidth}{X X X}
\toprule
\textbf{Benchmark} & \textbf{Features} & \textbf{Metrics} \\
\midrule
\rowcolor{gray!30} \multicolumn{3}{c}{\textbf{General Auditory Awareness and Processing}} \\
\midrule
\makecell[tl]{SALMon \\ \citep{salmonbench}} & Challenging benchmark for nuanced auditory awareness of semantic, paralinguistic, and acoustic information. & Likelihood-based Comparison \\
\midrule
\makecell[tl]{EmphAssess \\ \citep{seyssel2024emphassess}} & Benchmark testing awareness of prosodic features, requiring models to preserve them during speech-to-speech translation. & Precision, Recall, F1 \\
\midrule
\makecell[tl]{Dynamic-SUPERB \\ \citep{dynamicsuperb1}} & The first benchmark covering audio, speech, and music with 55 multiple-choice QA tasks. & Accuracy \\
\midrule
\makecell[tl]{Dynamic-SUPERB Phase-2 \\ \citep{dynamicsuperb2}} & An expanded version of Dynamic-SUPERB, currently the largest benchmark in the auditory processing category, featuring 180 tasks including open-ended ones. & Accuracy, TER, MSE, KTAU, LCC, SRCC, ERR, Miss Time, WER, Sacre BLEU, MER, IoU, F1, MEDAE, Angle Diff, Abs Diff, PER, PCC, DER, CER, POS \\
\midrule
\makecell[tl]{AIR-Bench \\ \citep{airbench}} & Benchmark covering audio, speech, and music with both multiple-choice and open-ended questions. & Accuracy, LLM-as-a-judge \\
\midrule
\makecell[tl]{AudioBench \\ \citep{audiobench}} & Benchmark focusing on speech and audio, comprising 8 tasks and 26 datasets. & Word Error Rate, LLM-as-a-judge, METEOR \\
\midrule
\makecell[tl]{MuChoMusic \\ \citep{muchomusic}} & In-depth investigation of music-oriented knowledge and processing abilities. & Accuracy, Instruction-following rate \\
\midrule
\rowcolor{gray!30} \multicolumn{3}{c}{\textbf{Knowledge and Reasoning}} \\
\midrule
\makecell[tl]{ZeroSpeech 2021 \\ \citep{zerospeech}} & Evaluating linguistic knowledge (lexical and syntactic understanding), primarily in English. & Likelihood-based Comparison, Similarity Comparison \\
\midrule
\makecell[tl]{CSZS \\ \citep{cszs}} & Assessing semantic and syntactic knowledge in code-switching scenarios (English-Mandarin, English-French, English-Spanish). & Likelihood-based Comparison \\
\midrule
\makecell[tl]{sStoryCloze \& tStoryCloze \\ \citep{storycloze}} & Benchmarks for semantic coherence in spoken story continuation. Focused on English. & Likelihood-based Comparison \\
\midrule
\makecell[tl]{MMAU \\ \citep{mmau}} & Expert knowledge and advanced reasoning (e.g., temporal, causal reasoning across audio, speech, music). & Accuracy \\
\midrule
\makecell[tl]{Audiopedia \\ \citep{audiopedia}} & Knowledge-intensive QA benchmark for evaluating world knowledge of entities. & Accuracy, F1 \\
\midrule
\makecell[tl]{VoiceBench \\ \citep{voicebench}} & Comprehensive benchmark for world knowledge, instruction following, safety, and robustness to acoustic variations. & LLM-as-a-judge, Accuracy, Refusal Rate \\
\midrule
\makecell[tl]{VoxEval \\ \citep{voxeval}} & Spoken version of MMLU considering speaker, style, and audio quality variations. & Accuracy \\
\midrule
\makecell[tl]{CompA \\ \citep{compa}} & Compositional reasoning of temporal order and attribute binding of sound events and captions. & Self-defined Text, Audio, and Group Scores \\
\midrule
\makecell[tl]{SAKURA \\ \citep{sakura}} & Multi-hop reasoning benchmark integrating knowledge and auditory information (gender, emotion, language, animal sounds). & Accuracy \\
\midrule
\makecell[tl]{URO-Bench \\ \citep{urobench}} & Benchmark of knowledge, safety, and instruction-following (English, Chinese, code-switching). & Word Error Rate, Character Error Rate, LLM-as-a-judge, UTMOS, Latency \\
\midrule
\makecell[tl]{Audio Entailment \\ \citep{audio_entailment}} & Deductive reasoning from auditory information. & Accuracy, Precision, Recall, F1 \\
\midrule
\citet{JASCO} & Joint reasoning integrating acoustic and speech information. & LLM-as-a-judge \\
\bottomrule
\end{tabularx}
\caption{Summary of the features and metrics of benchmarks in the \textbf{General Auditory Awareness and Processing} and the \textbf{Knowledge and Reasoning} categories in Figure~\ref{fig:taxonomy}.}
\label{tab:benchmarks_part1}
\end{table*}

\begin{table*}[t]
\centering
\small
\renewcommand{\arraystretch}{0.95}
\setlength\tabcolsep{4pt}
\begin{tabularx}{0.9\textwidth}{X X X}
\toprule
\textbf{Benchmark} & \textbf{Features} & \textbf{Metrics} \\
\midrule
\rowcolor{gray!30} \multicolumn{3}{c}{\textbf{Dialogue-oriented Ability}} \\
\midrule
\makecell[tl]{StyleTalk \\ \citep{styletalk}} &
Earliest benchmark for affective and contextual conversational abilities of LALMs. Primarily focuses on speaker emotion and speaking styles. &
BLEU, ROUGE, METEOR, BERT Score, F1, Human Evaluation (with A/B test) \\
\midrule
\makecell[tl]{SD-Eval \\ \citep{sdeval}} &
Benchmark for conversational abilities of LALMs, covering speaker emotion, accent, age, and environmental sounds. &
LLM-as-a-judge, ROUGE-L, BLEU, METEOR, BERT Score, Human Evaluation \\
\midrule
\makecell[tl]{VoxDialogue \\ \citep{voxdialogue}} &
Benchmark for conversational abilities of LALMs, further expanding the scope to 12 auditory attributes. &
BLEU, ROUGE-L, METEOR, BERT Score, F1, LLM-as-a-judge \\
\midrule
\makecell[tl]{Talking Turns \\ \citep{talkingturns}} &
Benchmark for evaluating the turn-taking dynamics of LALMs in full-duplex dialogues. &
Automatic judgment with an internally trained judge model \\
\midrule
\makecell[tl]{Full-Duplex-Bench \\ \citep{fullduplexbench}} &
Benchmark for pause handling, backchanneling, turn-taking, and interruption management in full-duplex dialogues. &
Takeover Rate, Backchannel Frequency, Jensen-Shannon Divergence, Latency, LLM-as-a-judge \\
\midrule
\makecell[tl]{Speech-IFEval \\ \citep{speech_ifeval}} &
Benchmark specifically tailored for instruction following, disentangling instruction-following from speech perception. Can also analyze the degree of catastrophic forgetting in LALMs compared with their LLM backbones. &
Word Error Rate, Accuracy, LLM-as-a-judge, Instruction-following Rate \\
\midrule
\rowcolor{gray!30} \multicolumn{3}{c}{\textbf{Fairness, Safety, and Trustworthiness}} \\
\midrule
\citet{lin2024listen} &
Quantifying the gender bias of LALMs through four tasks. &
Accuracy, F1, F1 Differences, Language Modeling Score, Stereotypical Score, Idealized Context Association Tests Score, Instruction Following Rate, Bias Score \\
\midrule
\makecell[tl]{Spoken Stereoset \\ \citep{lin2024spoken}} &
Evaluating social bias in LALMs, including gender and age. &
Speech Language Instruction Following Score, Speech Language Modeling Score, Speech Language Bias Score \\
\midrule
\citet{yang-etal-2025-audio} &
Quantifying safety issues of LALMs under jailbreaking attempts. &
Attack Success Rate by Attempt, Attack Success Rate by Questions \\
\midrule
\citet{multiaudiojail} &
Evaluating safety alignment of LALMs under adversarial multilingual and multi-accent audio jailbreaks. &
Jailbreak Success Rate, Word Error Rate, Accuracy \\
\midrule
\citet{kuan24_interspeech} &
Benchmark for object hallucination in LALMs, covering both discriminative and generative tasks. &
Accuracy, Precision, Recall, F1, Word Error Rate, Self-defined Metrics \\
\midrule
\makecell[tl]{CMM \\ \citep{cmm}} &
Benchmark for quantifying object hallucination and identifying its potential causes. Includes vision modality as well. &
Perception Accuracy, Hallucination Resistance \\
\bottomrule
\end{tabularx}
\caption{Summary of the features and metrics of benchmarks in the \textbf{Dialogue-oriented Ability} and the \textbf{Fairness, Safety, and Trustworthiness} categories in Figure~\ref{fig:taxonomy}.}
\label{tab:benchmarks_part2}
\end{table*}
\begin{table*}[!ht]
\setlength\tabcolsep{3pt}
\renewcommand{\arraystretch}{0.9}
\centering
\resizebox{0.95\linewidth}{!}{
\begin{tabular}{ccccccccc}
\toprule
\rowcolor{gray!30} \multicolumn{9}{c}{\textbf{General Auditory Awareness and Processing}} \\
\midrule
\multirow{2}{*}{\textbf{Benchmark}} & \multicolumn{4}{c}{\textbf{Input Modalities}} & \multicolumn{4}{c}{\textbf{Output Modalities}} \\
\cmidrule(lr){2-5} \cmidrule(lr){6-9}
& Text & Audio & Speech & Music & Text & Audio & Speech & Music \\
\midrule
SALMon~\cite{salmonbench} &  & \cmark & \cmark &  &  \multicolumn{4}{c}{\footnotesize \makecell[c]{Likelihood-based evaluation.\\No output modality.}}  \\
\midrule
\citet{wu2024just} & \cmark &  & \cmark &  & \cmark &  &  &  \\
\midrule

EmphAssess~\citep{seyssel2024emphassess} & &  & \cmark &  &  &  & \cmark & 
\\
\midrule
\citet{deshmukh2025adiff} & \cmark & \cmark & \cmark &  & \cmark &  &  &  \\
\midrule
Dynamic-SUPERB~\citep{dynamicsuperb1} & \cmark & \cmark & \cmark & \cmark & \cmark &  &  &  
\\
\midrule
Dynamic-SUPERB Phase-2~\citep{dynamicsuperb2} & \cmark & \cmark & \cmark & \cmark & \cmark &  &  &  \\
\midrule
AIR-Bench~\citep{airbench} & \cmark & \cmark & \cmark & \cmark & \cmark &  &  &  \\
\midrule
AudioBench~\citep{audiobench} & \cmark & \cmark & \cmark &  & \cmark &  &  &  \\
\midrule
MuChoMusic~\citep{muchomusic} & \cmark &  &  & \cmark & \cmark &  &  &  \\
\midrule
FinAudio~\citep{cao2025finaudio} & \cmark &  & \cmark &  & \cmark &  &  &  \\
\midrule
SAGI~\citep{roadmap} & \cmark & \cmark & \cmark & \cmark & \cmark &  &  &  \\
\midrule
MAE~\citep{mae} & \cmark & \cmark & \cmark &  & \cmark &  &  &  \\
\midrule
RUListening~\citep{rulistening} & \cmark &  &  & \cmark & \cmark &  &  &  \\
\midrule
OpenMU-Bench~\citep{openmubench} & \cmark &  &  & \cmark & \cmark &  &  &  \\
\midrule
\citet{wang2025advancing} & \cmark &  & \cmark &  & \cmark &  &  &  \\
\midrule

\citet{gong2024av} & \cmark & \cmark & \cmark & \cmark & \cmark &  &  &  \\
\midrule
Audio-FLAN~\citep{audioflan} & \cmark & \cmark & \cmark & \cmark & \cmark & \cmark & \cmark & \cmark \\
\midrule
QualiSpeech~\citep{qualispeech} & \cmark &  & \cmark &  & \cmark &  &  &  \\
\midrule

EvalSIFT~\citep{evalsift} & \cmark &  & \cmark &  & \cmark &  & \cmark &  \\
\midrule
OpenAQA~\citep{openaqa} & \cmark & \cmark &  &  & \cmark &  &  &  \\
\midrule

Clotho-AQA~\citep{clothoaqa} & \cmark & \cmark &  &  & \cmark &  &  &  \\
\midrule
SpeechCaps~\citep{speechcaps} & \cmark &  & \cmark &  & \cmark &  &  &  \\
\midrule
ASR-EC~\citep{asrec} & \cmark &  & \cmark &  & \cmark &  &  &  \\
\midrule
SLU-GLUE~\citep{slu-glue} & \cmark &  & \cmark &  & \cmark &  &  &  \\
\midrule
BEANS-Zero~\citep{beans-zero} & \cmark &  &  & \cmark & \cmark &  &  &
\\
\bottomrule
\end{tabular}
}
\caption{Input and output modalities of benchmarks in the \textbf{General Auditory Awareness and Processing} category shown in Figure~\ref{fig:entire_taxonomy}.}
\label{tab:modality_part1}
\end{table*}
\begin{table*}[ht]
\setlength\tabcolsep{4pt}
\renewcommand{\arraystretch}{0.9}
\centering
\resizebox{0.95\linewidth}{!}{
\begin{tabular}{ccccccccc}
\toprule
\rowcolor{gray!30} \multicolumn{9}{c}{\textbf{Knowledge and Reasoning}} \\
\midrule
\textbf{Benchmark} & \multicolumn{4}{c}{\textbf{Input Modalities}} & \multicolumn{4}{c}{\textbf{Output Modalities}} \\
\cmidrule(lr){2-5} \cmidrule(lr){6-9}
& Text & Audio & Speech & Music & Text & Audio & Speech & Music \\
\midrule

ZeroSpeech 2021~\citep{zerospeech} &  &  & \cmark &  &  \multicolumn{4}{c}{\footnotesize \makecell[c]{Likelihood-based evaluation.\\No output modality.}}  \\
\midrule

CSZS~\citep{cszs} &  &  & \cmark &  &  \multicolumn{4}{c}{\footnotesize \makecell[c]{Likelihood-based evaluation.\\No output modality.}}   \\
\midrule

sStoryCloze~\citep{twist} &  &  & \cmark &  & \multicolumn{4}{c}{\footnotesize \makecell[c]{Likelihood-based evaluation.\\No output modality.}}  \\
\midrule

tStoryCloze~\citep{twist} &  &  & \cmark &  & \multicolumn{4}{c}{\footnotesize \makecell[c]{Likelihood-based evaluation.\\No output modality.}}  \\
\midrule
BabySLM~\citep{babyslm} & \cmark &  & \cmark &  & \multicolumn{4}{c}{\footnotesize \makecell[c]{Likelihood-based evaluation.\\No output modality.}}  \\
\midrule


CompA~\citep{compa} & \cmark & \cmark &  &  &  \multicolumn{4}{c}{\footnotesize \makecell[c]{Similarity-based evaluation on\\text and audio inputs.}}  \\
\midrule

MMAU~\citep{mmau} & \cmark & \cmark & \cmark & \cmark & \cmark &  &  &  \\
\midrule
Audiopedia~\citep{audiopedia} & \cmark &  & \cmark &  & \cmark &  &  &  \\
\midrule
VoiceBench~\citep{voicebench} & \cmark &  & \cmark &  & \cmark &  &  &  \\
\midrule
VoxEval~\citep{voxeval} &  &  & \cmark &  &  &  & \cmark &  \\
\midrule

SAKURA~\citep{sakura} & \cmark & \cmark & \cmark &  & \cmark &  &  &  \\
\midrule
URO-Bench~\citep{urobench} & \cmark &  & \cmark &  & \cmark &  & \cmark &  \\
\midrule
Audio Entailment~\citep{audio_entailment} & \cmark & \cmark &  &  & \cmark &  &  &  \\
\midrule
ADU-Bench~\citep{adubench} &  &  & \cmark &  & \cmark &  & \cmark &  \\
\midrule
SAGI~\citep{roadmap} & \cmark & \cmark & \cmark & \cmark & \cmark &  &  &  \\
\midrule
MuChoMusic~\citep{muchomusic} & \cmark &  &  & \cmark & \cmark &  &  &  \\
\midrule
RUListening~\citep{rulistening} & \cmark &  &  & \cmark & \cmark &  &  &  \\
\midrule
OpenMU-Bench~\citep{openmubench} & \cmark &  &  & \cmark & \cmark &  &  &  \\
\midrule
\citet{gong2024av} & \cmark & \cmark & \cmark & \cmark & \cmark &  &  &  \\
\midrule
CompA-R~\citep{gama_compaR} & \cmark & \cmark &  &  & \cmark &  &  &  \\
\midrule
OpenAQA~\citep{openaqa} & \cmark & \cmark &  &  & \cmark &  &  &  \\
\midrule
Clotho-AQA~\citep{clothoaqa} & \cmark & \cmark &  &  & \cmark &  &  & \\
\midrule
SLU-GLUE~\citep{slu-glue} & \cmark &  & \cmark &  & \cmark &  &  &  \\
\midrule
SpeechCaps~\citep{speechcaps} & \cmark &  & \cmark &  & \cmark &  &  &  \\
\midrule
\citet{JASCO} & \cmark & \cmark & \cmark &  & \cmark &  &  &  \\
\midrule
\citet{deshmukh2025adiff} & \cmark & \cmark & \cmark &  & \cmark &  &  &  \\
\bottomrule
\end{tabular}
}

\caption{Input and output modalities of benchmarks in the \textbf{Knowledge and Reasoning} category shown in Figure~\ref{fig:entire_taxonomy}.}
\label{tab:modality_part2}
\end{table*}
\begin{table*}[ht]
\setlength\tabcolsep{3pt}
\renewcommand{\arraystretch}{0.9}
\centering
\resizebox{0.95\linewidth}{!}{
\begin{tabular}{ccccccccc}
\toprule
\rowcolor{gray!30} \multicolumn{9}{c}{\textbf{Dialogue-oriented Ability}} \\
\midrule
\textbf{Benchmark} & \multicolumn{4}{c}{\textbf{Input Modalities}} & \multicolumn{4}{c}{\textbf{Output Modalities}} \\
\cmidrule(lr){2-5} \cmidrule(lr){6-9}
& Text & Audio & Speech & Music & Text & Audio & Speech & Music \\
\midrule

 StyleTalk~\citep{styletalk} & \cmark & & \cmark & & & & \cmark & \\
\midrule
SD-Eval~\citep{sdeval} & \cmark & \cmark & \cmark & & \cmark & & & \\
\midrule
VoxDialogue~\citep{voxdialogue} &  & \cmark & \cmark & \cmark & \cmark & & & \\

\midrule
Talking Turns~\citep{talkingturns} & & & \cmark & & & & \cmark & \\
\midrule
Full-Duplex-Bench~\citep{fullduplexbench} & & & \cmark & & & & \cmark & \\
\midrule
\citet{mindthegap} & & & \cmark &  & \cmark & &  & \\
\midrule
ContextDialog~\citep{contextdialog} & & & \cmark & & \cmark & & \cmark & \\
\midrule
ADU-Bench~\citep{adubench} &  &  & \cmark &  & \cmark &  & \cmark &  \\
\midrule
VoiceBench~\citep{voicebench} & \cmark & & \cmark & & \cmark & & & \\
\midrule
URO-Bench~\citep{urobench} & \cmark & & \cmark & & \cmark & & \cmark & \\
\midrule
Speech-IFeval~\citep{speech_ifeval} & \cmark & & \cmark & & \cmark & & & \\
\midrule
S2S-Arena~\citep{s2sarena} & & & \cmark & & & & \cmark & \\
\midrule
EvalSIFT~\citep{evalsift} & \cmark &  & \cmark &  & \cmark &  & \cmark &  \\
\bottomrule
\end{tabular}
}

\caption{Input and output modalities of benchmarks in the \textbf{Dialogue-oriented Ability} category shown in Figure~\ref{fig:entire_taxonomy}.}
\label{tab:modality_part3}
\end{table*}
\begin{table*}[ht]
\setlength\tabcolsep{3pt}
\renewcommand{\arraystretch}{0.9}
\centering
\resizebox{0.95\linewidth}{!}{
\begin{tabular}{ccccccccc}
\toprule
\rowcolor{gray!30} \multicolumn{9}{c}{\textbf{Fairness, Safety, and Trustworthiness}} \\
\midrule
\textbf{Benchmark} & \multicolumn{4}{c}{\textbf{Input Modalities}} & \multicolumn{4}{c}{\textbf{Output Modalities}} \\
\cmidrule(lr){2-5} \cmidrule(lr){6-9}
& Text & Audio & Speech & Music & Text & Audio & Speech & Music \\
\midrule

\citet{lin2024listen} & \cmark & & \cmark & & \cmark & & & \\
\midrule
Spoken Stereoset~\citep{lin2024spoken} & \cmark & & \cmark & & \cmark & & & \\
\midrule
VoiceBench~\citep{voicebench} & \cmark & & \cmark & & \cmark & & & \\
\midrule
\citet{yang-etal-2025-audio} & \cmark & \cmark & \cmark & & \cmark & & & \\
\midrule
\citet{multiaudiojail} & \cmark & & \cmark & & \cmark & & & \\
\midrule
AdvBench-Audio~\citep{advbench_audio} & \cmark & & \cmark & & \cmark & & & \\
\midrule
\citet{xiao2025tune} & \cmark & & \cmark & & \cmark & & & \\
\midrule
\citet{gupta2025bad} & \cmark & & \cmark & & \cmark & & & \\
\midrule
\citet{hughes2024best} & \cmark & & \cmark & & \cmark & & & \\
\midrule
URO-Bench~\citep{urobench} & \cmark &  & \cmark &  & \cmark &  & \cmark &  \\
\midrule
\citet{kuan24_interspeech} & \cmark & \cmark & \cmark & & \cmark & & & \\
\midrule
CMM~\citep{cmm} & \cmark & \cmark & & & \cmark & & & \\
\midrule
\citet{match} & \cmark & \cmark & & & \cmark & & & \\

\bottomrule
\end{tabular}
}

\caption{Input and output modalities of benchmarks in the \textbf{Fairness, Safety, and Trustworthiness} category shown in Figure~\ref{fig:entire_taxonomy}.}
\label{tab:modality_part4}
\end{table*}



\end{document}